\documentclass[twocolumn,aps,prxlife]{revtex4-2} \usepackage[T1,T2A]{fontenc} \usepackage{graphicx} \usepackage{amsmath}  \usepackage{hyperref} \usepackage{amssymb}\usepackage{amsfonts} \usepackage{tikz} \usepackage{calc} \newlength{\kwidth}  \def\mi{\textrm{i}\hspace{0.2ex}} \def\me{\textrm{e}} \def\md{\textrm{d}} \def\tr{\mbox{tr}} \def\kT{k\hspace{-0.3\kwidth}T} \usepackage{verbatim} \usepackage{xcolor}

\begin{document}

\title{Quantum Dynamics of a Nanorotor Driven by a Magnetic Field }

\author{V. N. Binhi} \email{vnbin@mail.ru}

\begin{abstract} A molecular rotor mechanism is proposed to explain weak magnetic field effects in biology. Despite being nanoscale (1\,nm), this rotor exhibits quantum superposition and interference. Analytical modeling shows its quantum dynamics are highly sensitive to weak, but not strong, magnetic fields. Due to its enhanced moment of inertia, the rotor maintains quantum coherence relatively long, even in a noisy cellular environment. Operating at the mesoscopic boundary between quantum and classical behavior, such a rotor embedded in cyclical biological processes could exert significant and observable biological influence.

\end{abstract}

\maketitle

\section{Introduction}

The spin-chemical radical pair mechanism (RPM) is the leading theory today for explaining the biological effects of low-intensity magnetic fields \citep{Buchachenko-2014-2, Hore-and-Mouritsen-2016, Binhi-2019}. However, the reason remains unclear why the observed magnetic {biological} effects often have a large relative magnitude, ranging from units to tens of percent, while the spin-{chemical} RPM effects with cryptochromes in vitro are typically less than a tenth of a percent.

RPM effects arise when the quantum system of radicals is well-protected from the perturbing influence of the environment. In other words, these effects are closely related to the spin coherence relaxation rate $g$.

The analytical solution of the RPM master equation, the Liouville–von Neumann equation with decoherence and chemical terms, \begin{equation} \label{master} \partial_t \, \rho = -\mi \left[ {\cal H} ,\rho \right]  - k \rho  - g \left[\rho- \tr (\rho) \bar{\rho} \right] \end{equation} for the density matrix $\rho$ of an idealized system of two electrons and one nucleus with minimal interactions, in a static MF was obtained in \citep{Binhi-2025-PRE}. A relaxation superoperator of the form $ -g[\rho - \tr (\rho ) \bar{\rho}]$ was used, where $\bar{\rho} $ is the final density matrix corresponding to thermal equilibrium. The case of equal relative rates of chemical kinetics for the singlet and triplet channels $k_{\textrm{S}}=k_{\textrm{T}}=k$ was considered.

A solution of this equation for realistic values of the rates $k$ and $g$ shows only a very slight change in $\rho$ in weak MFs. This fails to account for many magnetobiological effects, particularly the effects of magnetic storms and the ability of many animals to utilize the geomagnetic landscape for their seasonal migrations.

It appears that, while remaining within the framework of physics, it is impossible to overcome the small magnitude of the primary RPM magnetic response. However, organisms seem to somehow manage to either comply with this constraint or circumvent it. Consequently, magnetoreception mechanisms distinct from the RPM must be considered---particularly those reliant on complex biological processes that elude in vitro modeling. Such processes include those with a recurring catalytic cycle, such as DNA/RNA biopolymerization, protein synthesis, and folding. The probability of an error in these (quasi)cyclic processes increases with the length of the synthesized or folded chain and may depend on the MF \citep{Binhi-2023-b}.

On one hand, in these processes, large molecular fragments undergo diverse rotations, thus representing nanoscopic rotors. On the other hand, even in neutral fragments, the spatial distributions of negative and positive electric charges do not coincide. This occurs due to the displacement of electron clouds from their nuclei, which is confirmed by the large electric dipole moments of the fragments. During rotations, the electric charges generate an MF, or acquire magnetic moment, that interacts with the external MF.

For instance, during protein synthesis and folding, rotations of amino acid residues occur. Since the residue has a distributed charge, its rotation induces a magnetic moment, the dynamics of which in an MF are described in terms of the gyromagnetic ratio $\gamma$. The thermalization time $\tau$ of the rotational state for the residue can reach up to 0.1\,s \citep{Binhi-and-Savin-2002}.

An amino acid residue is a massive molecular construct compared to a single atom. If it is in a thermalized state, the most effective description is in terms of classical rotational dynamics. However, due to the relative slowness of thermalization after the birth or release of the rotor, the emergence of quantum effects is probable. This approach is adopted in the present study. The aim was to evaluate the specific quantum rotation of a nanorotor arising from the interference of its quantum states. Such rotation could provide an explanation for the biological effects of weak MFs, offering an alternative to the RPM.

\section{Quantum rotor in an MF} \label{qrdynamics}

The formal solution of Eq.~(\ref{master}) has the form \citep{Binhi-2025-PRE} \begin{align} \label{solut} \rho (t) = \theta_1 \, U \rho_0 U^{\dagger}  +  \theta_2 \bar{\rho}, \\ \theta_1 \equiv  \me ^{-jt}, ~~ \theta_2 \equiv  \me ^ {- kt} \left(1 - \me^ {-gt} \right) , \nonumber \end{align} where $U(t) \equiv \exp( -\textrm{i} {\cal H}t)$ is the Hamiltonian evolution operator. Applied to a rotor, whose Hilbert state space is infinite-dimensional, it is convenient to switch to a finite-dimensional basis in which the equilibrium density matrix is described by a Boltzmann distribution. In this case, the completely mixed state with a diagonal, equally populated density matrix will be a possible approximation.

We use a basis of dimension $2L+1$, where $L$ is the maximum value of the orbital quantum number, or angular momentum, sufficiently large to populate rotor states with energy up to $\kT$ under the action of thermal perturbations. Population of states with larger $L$ is unlikely. The idealization for analytical calculations is that thermal perturbations (in the absence of chemical kinetics) cause a redistribution of populations from the initial state $\rho_0$ to the thermal equilibrium state $\bar{\rho}$ with a Boltzmann distribution.

In the quantum rotor model, as a rough approximation, we assume that the chemical process causes a reduction of $\rho$ from all states, and then the chemical superoperator has the same form $-k\rho$ as in the RPM. For this case, solution (\ref{solut}) is valid, and the density matrix undergoes a ``chemical'' reduction of the form $\tr(\rho)= \me^{-kt}$.

The idealization of a molecular rotor is a rigid system of masses and charges rotating about a fixed axis. The Hamiltonian of such a rotor in a constant uniform MF $H$ along the $z$ axis, was obtained in \citep{Binhi-2002} in the form $  {\cal H} = \Lambda^2 /{2I}\equiv  \left( \hbar {\cal L} - I \gamma H \right)^2 /{2I}, $ where $\Lambda \equiv \hbar {\cal L} - I \gamma H$ is the kinetic (measurable) angular momentum, ${\cal L} \equiv - \mi \partial / \partial\varphi$ is the dimensionless operator of the canonical angular momentum, $I$ is the moment of inertia and $\gamma$ is the orbital gyromagnetic ratio of the molecule. Further, the Hamiltonian term proportional to $H^2$ will be omitted due to its smallness.

For convenience, we introduce the unit of frequency $a\equiv \hbar/2I$ and dimensionless parameters of time, MF, rates and energy according to the following relations, \begin{multline} \label{dimless} t\equiv at',~~x\equiv 2\gamma H I/\hbar,~~k\equiv \kappa/a, \\ g\equiv \Gamma/a \equiv 1/a\tau, ~~ \varepsilon = E/\hbar a,  \end{multline} where $t'$ is time, $x$ is proportional to the MF, $\kappa$ and $\Gamma = 1/\tau$ are the physical rates of chemical kinetics and thermal relaxation, $\tau$ is the relaxation time of quantum coherence, $E$ is eigenenergy. Then the Hamiltonian is \begin{equation} \label{hamiltonian} {\cal H} = {\cal L}^2 - x{\cal L} , \end{equation} and the Schrödinger equation reduces to the dimensionless form $\partial_t \Psi(\varphi ) = -\mi {\cal H} \Psi(\varphi )$. For nanorotors $x\ll 1$.

The eigenvectors $|n\rangle$ of the Hamiltonian (\ref{hamiltonian}) in the energy representation correspond to the orthonormal functions of the coordinate representation $\phi_n (\varphi)$ and the eigenvalues $\varepsilon_n$ \begin{equation} \label{eigen} \phi_n = \frac1{\sqrt {2\pi}} \exp(\mi n\varphi),~ \varepsilon_n = n^2 - xn, ~ n=-L, ..., L \end{equation} where $n$ is the magnetic, or orbital, quantum number, $L$ is its maximum value, or maximum angular momentum, and $\varphi$ is the angular coordinate.

Further, the Boltzmann distribution $\exp (-E_n/\kT ) /Z$ will be needed, where $E_n = \hbar^2 (n^2 - xn)/2I$ are the eigenenergies of the rotor and $Z$ is the partition function. In dimensionless form, the distribution is \begin{equation} \label{vardefs} R_n \equiv \frac 1Z \me ^ {- \beta n^2 }, ~~ Z \equiv \sum_n \me ^ {- \beta n^2 }, ~~\beta \equiv \frac {\hbar^2} {2 I \kT} \end{equation} where the small addition $xn$ is neglected. Here and below, sums are taken from $-L$ to $L$. The condition $ (\hbar^2 / 2I)\varepsilon_n \sim \kT$ means thermal population of the rotor states and determines $n \approx 94$. With some margin, so as to account for the majority of populated states, we set $L = 200$ below. To explain how the magnetic effect arises, one must consider the processes in the coordinate representation rather than the energy representation. Note that the average angular momentum of the rotor is $ \tr (\Lambda \rho)$, where $\Lambda \equiv \hbar {\cal L} - I \gamma H = \hbar {\cal L} - \hbar x/2 $ is the operator of the kinetic angular momentum. From solution (\ref{solut}) it can be shown that for a rotor with a pure initial state $\tr({\cal L}\rho)=0$, and consequently, the angular momentum is $-\tr(I\gamma H \rho) = -I\gamma H \me^{-kt} = - (\hbar /2)x \, \me^{-kt}$, and the magnetic moment is $-I\gamma^2 H \me^{-kt}$. It can be seen that the moment is of a diamagnetic nature, i.e., it is induced by the MF and directed opposite to the field. Magnetic resonances cannot occur in this system because the magnetic moment of the rotor is induced, not intrinsic. Moreover, it is extremely small due to the proportionality to $x$ ($x\ll 1$), and plays no role in the energetics of quantum rotations. However, the motion of the wave functions of the rotor's probability density not only creates this moment but also constitutes the basis of an important interference effect, which becomes visible in the coordinate representation.

\begin{figure} [htbp] \centering \includegraphics[width=0.82\linewidth]{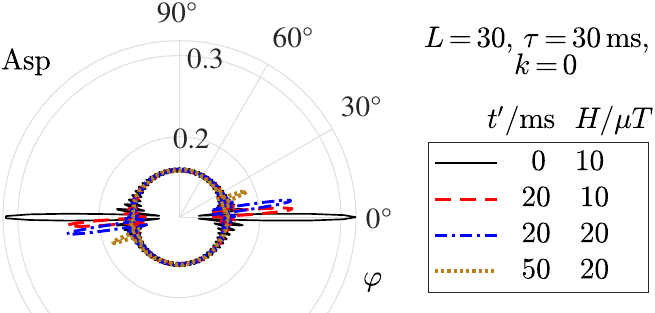} \caption{Rotation of the probability density $W$ of the rotor, representing the Asp residue, at different values of physical time and MF and in the absence, for clarity, of chemical reduction. The probability density features a prominent narrow spike.} \label{fig:rotorpozition} \end{figure}

The density $W$ of the probability of finding the rotor at position $\varphi$ is the ``diagonal'' element of the density matrix in the coordinate representation, \[ W(t,\varphi) =  \sum_{nm}  \rho_{nm} (t) \, \phi_n^*(\varphi) \phi_m (\varphi)  . \] From definitions (\ref{solut}) and (\ref{eigen}) follows the normalization of the density matrix to its chemically reduced value: $\int _0^{2\pi} W(\varphi, t) \md \varphi =   \sum_{nm}  \rho_{nm} \delta_{nm} = \tr(\rho) = \me^{-kt}$. Substituting $\rho$ and $\phi $ from these definitions, we write: \begin{equation} \label{probdensity} W(t,\varphi) = \frac 1{2\pi} \left(\theta_1 s_1 + \theta_2  s_2 \right), \end{equation} where it is taken into account that $\bar{\rho}_{nm}=\delta_{nm}/N$ ($\beta=0$ for simplicity), where $N\equiv 2L+1$ and notations for the sums $s$ are introduced: \begin{multline} \label {sdefs} s_1 \equiv \sum_{nm} \me ^{-\mi (n-m)\varphi} \left( U \rho_0 U^{\dagger} \right) _{nm}, \\ s_2 \equiv \sum_{nm} \me ^{-\mi (n-m)\varphi} \,\delta_{nm} \end{multline} To calculate the sum $s_1$, let us first find the elements of the matrix $U \rho_0 U^{\dagger}$, where we set $\rho_0 = |\psi \rangle \langle \psi |$, and immediately write the initial state for the case $\beta=0$, i.e. $|\psi \rangle = (1/\sqrt{N}) \sum_i |i\rangle $. Then, applying the spectral decomposition of $U$ and performing calculations, we get $ U \rho_0 U^{\dagger} = \sum_i \pi_i \lambda_i |\psi \rangle \langle \psi | \sum_k \pi_k \lambda_k^* = \lambda_n \lambda_m^* /N$, i.e. $ \left( U \rho_0 U^{\dagger} \right)_{nm} =  \lambda_n \lambda_m^*/N $. Substituting this into $s_1$ and recalling that $\lambda_n \equiv \exp(-\mi \varepsilon_n t) =\exp\left[\mi (nx-n^2) t \right]$, we obtain the sum in the form \[ s_1 = \frac1N \sum_{nm} \me ^{-\mi (n-m)(xt-\varphi )} \me ^{ -\mi (n^2-m^2) t} . \] Note that all sums are taken from $-L$ to $L$. In this sum there are terms with $m=\pm n$, and extremely rapidly oscillating terms with $m\neq \pm n$ with about $N^2/2$ different frequencies. Therefore, the sum of these mutually compensating terms is very small and can be neglected. Calculating the terms with $m=\pm n$, we find \[ s_1 = 1 +\frac 1{N} \frac {\sin [ N (xt-\varphi)] } {\sin (xt-\varphi)} . \] The sum $s_2$ is, obviously, $N$. Hence, substituting $\theta$ and the found $s$ into (\ref{probdensity}) we obtain for $L\gg 1$ \begin{equation} \label{finprdensity} W(t,x,\varphi) = \frac {\me ^{-kt}} {2\pi} \left[ 1 +  { \me ^{-gt}}  \, \frac {\sin [ 2L (xt-\varphi)] } {{2L} \sin (xt-\varphi)}  \right]  . \end{equation} Figure~\ref{fig:rotorpozition} shows this relation in the form of a $\pi$-periodic function of the polar angle $\varphi$ for different values of time and MF, which were converted from the dimensionless parameters $t$ and $x$ according to (\ref{dimless}). Calculations are performed for a nanoscopic rotor in the form of an aspartic acid (Asp) residue. Estimation of parameter values is given in the next section.

It can be seen from the figure that the probability density of the rotor being at the angular position $\varphi$ near the initial position $xt-\varphi =0$ is a wave packet having a narrow spike with a maximum height of $1/\pi$ against a constant background of slightly less than $1/2\pi$. Its half-width is $1/L$. The nanoscopic nature of the rotor is manifested here in the narrowness of the spike. This feature represents a veritable quantum needle, which ``probes'' its immediate environment. As the mass of the rotor increases, the number of populated states increases and their interference produces an increasingly narrow spike. It disappears in the macroscopic limit. At the same time, the constant component tends to $1/2\pi$.

This is how the transition from quantum to classical dynamics occurs here. It can also be said differently: the peak of the wave packet rotates, as follows from the argument $xt-\varphi$, with an angular velocity $x$, or $\gamma H$ in physical units. The greater the MF, the faster this ``diamagnetic'' rotation. The direction of rotation is determined by the sign of the MF, and its speed is proportional to $\gamma$. As the size of the rotor increases, the gyromagnetic ratio decreases and this effect disappears.

The transition from quantum to classical motion also occurs in time: as the rotor thermalizes, coherent rotation is replaced by Brownian diffusive rotation. The latter is represented phenomenologically by the coefficient $g$. With time $t$ the needle in Fig.\,\ref{fig:rotorpozition} shifts; simultaneously its height, due to the factor $\me ^ {-(k+g)t }$ in (\ref{finprdensity}), decreases rapidly. The shift by half the peak width occurs in time $t = 1/xL$. For the magnetic effect to be noticeable, the decrease in height should not be too fast, i.e. $(k+g)t < 1$, or $xL > k+g$. In terms of physical variables, see (\ref{dimless}), this inequality determines the minimal detectable MF \begin{equation} \label{fundlim} \gamma H \tau \sim \frac 1L (1+\kappa\tau ), \end{equation} which implies that the nanorotor's sensitivity to the MF is $L$ times higher than that in the RPM \citep{Binhi-2025-UFN-e}.

Let us recall that in this scenario, rotation in the geoMF, which displaces the peak away from its initial position where the probability of an erroneous biochemical reaction is increased, is the evolutionary norm. An adverse biological effect occurs if such rotation is absent, i.e., with a decrease in the MF to hypomagnetic levels \citep{Kaspranski-ea-2025}.

\section{Gyromagnetic ratio of the Asp amino acid residue} \label{estimateAsp}

The gyromagnetic ratio of the electron spin is $\gamma_{\text{sp}} = \mu_\me /\frac12 \hbar \approx 1.76\times 10^7$~rad\,G$^{-1}$s$^{-1}$. For the orbital motion of an electron, its gyromagnetic ratio is $ \gamma_{\text{or}} = e/ 2m _\me c \approx 8.79 \times 10^6$~rad\,G$^{-1}$s$^{-1}$, where $c$ is the speed of light in vacuum. It is slightly smaller than the spin one. Let us estimate the gyromagnetic ratio of an amino acid residue, assuming, as stated above, that various atoms and regions of the electric charge density of the molecule perform rotations, or orbital motions, creating the total mechanical and magnetic moments of the molecule.

The gyromagnetic factor $\gamma $ of the rotational motion of a system of point masses $m_i$ with charges $q_i$ is obtained from the gyromagnetic ratio of the orbital motion of an electron by the substitutions $m_{\textrm{e}} \rightarrow I$ and $e\rightarrow Q$, i.e. $\gamma \equiv Q/2Ic$, where $I\equiv \sum_i m_i r_i^2$ and $Q\equiv \sum_i q_i r_i^2$ are the moment of inertia of the masses and a characteristic of the charge distribution of the molecule, $r_i$ is the distance from the point mass $m_i$ and charge $q_i$ center to the axis of rotation. Amino acids with a dipole moment along the side chain obviously have the largest gyromagnetic ratios. Lysine and arginine have long aliphatic chains with charged groups at the end. In glutamic and aspartic acids, carboxyl groups create a significant dipole moment. Methionine and cysteine include sulfur-containing groups with high bond polarity. The maximum gyromagnetic ratio is for aspartic acid, Asp, due to the large charge of the hydroxyl oxygen, about 0.6\,$e$, at a distance of about 0.4~nm from the axis of rotation. Estimation taking into account the distribution of atomic charges and masses of the Asp residue gives $\gamma \approx 40$~rad\,G$^{-1}$s$^{-1}$, which is $10^5$ times smaller than electron $\gamma_{\text{or}}$. However, the decoherence time $\tau$ of the rotational state of such a residue can be 1--100~ms \citep{Binhi-and-Savin-2002}, which is approximately six orders of magnitude greater than the decoherence time of magnetosensitive electron radicals in proteins, 3--30~ns \citep{Binhi-2025-PRE}. Therefore, the product $\gamma\tau$, which determines the necessary minimum MF (\ref{fundlim}), can be even larger than that of the electron of a radical pair.

The angular velocity of rotation of the neutral form of the Asp residue in the geoMF is about $\gamma H_{\rm geo} \approx 20$~rad/s or about 1~deg/ms. Over the decoherence time of rotation, the angular displacement of the residue can amount to several degrees. From the standpoint of the structural geometry of the protein and the characteristic times of its folding, this value is significant. It is not so small that it can be neglected, but also not so large that one can assume a uniform averaged distribution of angular positions. If the magnitude of the displacement is comparable to the narrowness of the rotor's quantum needle, Fig.\,\ref{fig:rotorpozition}, then the absence of displacement initiates a biopolymerization error.

\section{Discussion} \label{discus}

In the illustrative vector model of spin, the RPM emphasizes the orientation of the spin magnetic moments of the radical pair relative to each other. The quantum rotor scenario develops the theme of peculiarities in the dynamics of a single abstract magnetic moment relative to its immediate environment \citep{Binhi-and-Savin-2002, Binhi-and-Prato-2018}.

In the present work, the scenario of the involvement of nanorotors in the biological response to weak MF variations receives significant development. We propose that during biological evolution, some processes with recurring catalytic cycles, particularly ribosomal protein synthesis, adapted to the presence of the geoMF. During the thermalization time, the geoMF rotates the rotor by a few degrees immediately after its release in a pure quantum state. This rotation is the norm, ensuring optimal synthesis and correct protein folding. If this rotation does not occur---for example, in the absence of an MF---the radical amino acid residues persist in a non-optimal position. This increases the probability of misincorporation and erroneous protein folding. The quantum rotor mechanism suggests that the organism is protected by the geoMF due to evolutionary adaptation. A reduction to hypoMF levels removes this protection, causing increased errors in biopolymerization and associated negative outcomes.

Interestingly, the width of the quantum needle in the rotor's probability density, Fig.\,\ref{fig:rotorpozition}, corresponds to the angular size of the sector (a few degrees) where the probability of an adverse biochemical reaction is increased. This size obviously could not be an order of magnitude smaller or larger. It is surprising that the number $L\sim 100$--$200$ of thermally excited eigenstates of the rotor determines the spike width corresponding to a reasonable sector size. Finally, it is also surprising that the temperature factor in this scenario not only limits the magnitude of the effect via decoherence but also determines the very possibility of a magnetic effect. If thermal population of the rotor states were absent, the interference pattern would not have a sharp spike, whose movement under the action of the MF is associated with the magnetic effect.

The fact that the quantum rotor mechanism considers quantum rotations does not at all mean that free full-circle (or more) rotation of the rotor is necessary. For the effect to occur, as we have seen, only minor angular displacements of the rotor from its initial position upon its formation are sufficient. Therefore, the fact that rotations of amino acid residues, for example, are limited by various steric factors or occur in complex rotational potentials, does not prevent the manifestation of the magnetic effect.

Above, we considered an example with an amino acid residue appearing in the active site of the ribosome during translation. Currently, no reasons are seen why similar processes could not occur in DNA--RNA biopolymerization enzymes, particularly in enzymes of the repair system.

Accounting for the intrinsic rotations of enzymes carrying molecular rotors ensures high sensitivity to MF variations against the background of the geoMF \citep{Binhi-and-Prato-2018}. This allows interpreting animal magnetic navigation, i.e., their use of the Earth's magnetic relief (variations at the level of tens of nT) in multi-thousand-kilometer migrations to seasonal habitats.

This could also explain the intriguing correlation between geomagnetic disturbances and some biological processes, such as those observed in birds under laboratory conditions \citep{Engels-ea-2014}. Evidence for a causal relationship in humans has long remained unconvincing, due to the disparity between the amplitudes of geomagnetic disturbances and those of the diverse MFs in which humans are constantly immersed \citep{Breus-ea-2016-e}. However, systems capable of recording and reproducing a magnetic storm in the laboratory have appeared, with interesting results on the direct action of artificial storms on humans \citep{Pishchalnikov-ea-2019}. In light of the quantum rotor mechanism presented above, these results do not appear contradictory.

\section{Conclusion} \label{conclud}

We propose a mechanism for the influence of weak MFs on biological systems, based on the quantum behavior of a nanoscopic molecular rotor. Despite thermal noise, this rotor exhibits long enough fine-structured interference sensitive to weak fields, offering a potential explanation for the effects of hypomagnetic fields and magnetic storms, and for animal magnetoreception.

The molecular rotor is not just a rotor obeying the laws of quantum physics, but a nanoscopic system whose typically quantum behavior is observable and significant under biological conditions. This is one of the cases of preserving quantum effects in biological systems \citep{Binhi-and-Rubin-2023e}. The peculiarities of nanoscopic dynamics, leading to an extreme responsiveness of the quantum needle to parameter variations, to MF variations in particular, represent a previously unstudied resource of biology. This opens a new perspective for theoretical and experimental research.

 \end{document}